\documentclass{article}
\usepackage{frascatiphys}
\usepackage{graphicx}
\usepackage{amsmath,amssymb}
\usepackage{url}

\newcommand{\ee}{e^-e^+}
\newcommand{\mw}{M_W}
\newcommand{\mev}{\mathrm{MeV}}
\newcommand{\gev}{\mathrm{GeV}}
\newcommand{\tev}{\mathrm{TeV}}

\begin{document}
\title{ 
Theory status of four-fermion production at $\ee$ colliders
}
\author{
Christian Schwinn\thanks{\,
 Heisenberg Fellow of the German Research Foundation (DFG).
}        \\
{\em Institute for Theoretical Particle Physics and Cosmology,}\\
 {\em     RWTH Aachen University, D-52056 Aachen, Germany}\\
 and \\
{\em Albert-Ludwigs-Universit\"at Freiburg, Physikalisches Institut,}\\ 
\em{  D-79104 Freiburg, Germany 
}
}
\maketitle
\baselineskip=11.6pt
\begin{abstract}
The  status of predictions for  four-fermion production at  $\ee$ colliders is
reviewed with an emphasis on the developments after the LEP2 era and
an outlook to the challenges posed by the precision program at future colliders.
\end{abstract}
\baselineskip=14pt

\section{Introduction}

After the discovery of a Higgs boson, the search for physics beyond the
Standard Model~(SM) of particle physics is one of the main objectives of run 2
of the LHC and of future colliders.  In case new particles are not directly
accessible at these colliders or in non-collider experiments, one can search
for indirect evidence for new physics through precise studies of
electroweak~(EW) or flavour observables, and the couplings of the gauge bosons
and the Higgs boson.  Further, accurate measurements of input parameters of
the SM such as the masses of the $W$ and $Z$ bosons and the top quark are
required for the precision-physics program.  Here future $\ee$ colliders could
play a particularly important role by revisiting the LEP precision
measurements at higher statistics, and further measuring top-quark and
Higgs-boson properties.  Currently linear colliders such as ILC and CLIC as
well as circular colliders such as FCC-ee or CEPS are
investigated.\cite{Baer:2013cma}

\begin{figure}[t]
    \begin{center}
       \includegraphics[width=.7\textwidth]{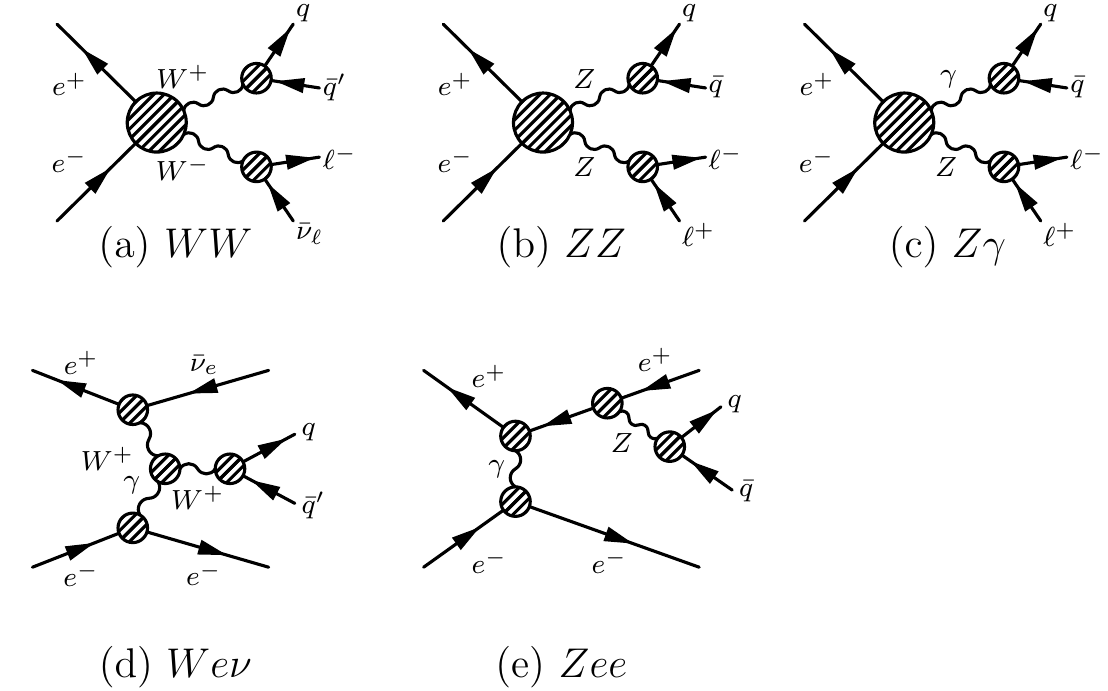}
       \caption{
         Classification of signatures in four-fermion production.
       }
\label{fig:processes}
    \end{center}
\end{figure}
An important signature at high-energy $\ee$ colliders is given by
\emph{four-fermion production processes}\footnote{Four-fermion final
  states arising from Higgs-boson production
  with subsequent decay to $b$ quarks or $\tau$ leptons are not considered in
  this contribution.}
as shown in Figure~\ref{fig:processes}.
They have been explored at
LEP2\cite{Schael:2013ita} 
for centre-of-mass energies $\sqrt s=161.3$-$206.6\,\gev$, allowing precision
tests of the SM through measurements of cross-sections, the mass, width and
branching ratios of the $W$-boson in $W$-pair production
(Fig~\ref{fig:processes} (a)), and triple-vector boson couplings in $W$-pair
production, $Z \gamma$ and single-$W$ production (Fig~\ref{fig:processes} (c)
and (d), respectively). At future $\ee$ colliders the precision of these
measurements could be increased, for instance by up to two magnitudes for the
triple gauge boson couplings.\cite{Baak:2013fwa} For $M_W$, an accuracy of
$3$--$4\,\mev$ is projected for an ILC, while $1\,\mev$ may be
possible using a threshold scan of the $W$-pair production cross section at a
future circular $\ee$ collider.\cite{Baak:2013fwa}

In this contribution, the theoretical
challenges and the  methods used for  four-fermion
production are discussed in Section~\ref{sec:theory}.
Recent theoretical results are reviewed in Section~\ref{sec:recent} while
an outlook to future developments  needed to meet the requirements of planned
colliders is given in Section~\ref{sec:outlook}.

\section{Theoretical challenges and methods}
\label{sec:theory}

\begin{figure}[t]
    \begin{center}
       \includegraphics[width=.65\textwidth]{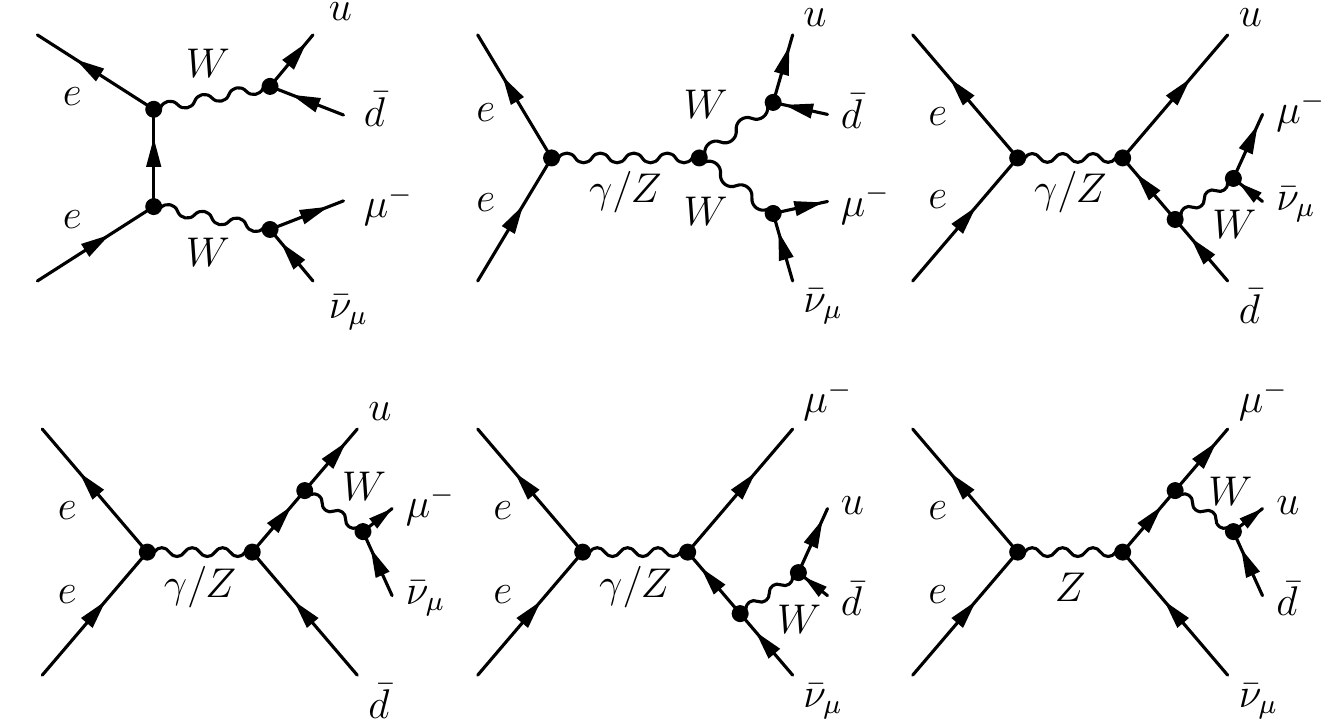}
        \caption{Diagrams contributing at tree-level to the $e^-e^+\to u\bar d
        \mu^-\bar\nu_\mu$ process.}
\label{fig:cc10}
    \end{center}
\end{figure}

In the theoretical description of four-fermion production, in general all
diagrams contributing to a given final state must be taken into
account for a consistent, gauge invariant result,\footnote{In some cases,
  gauge invariant subsets of diagrams can be
  identified.\cite{Grunewald:2000ju}} resulting in a large number of
contributing Feynman diagrams, in particular beyond leading order.  These
typically include topologies different from the resonant ``signal'' diagrams
of the  processes in Figure~\ref{fig:processes}.  For
instance, as shown in Figure~\ref{fig:cc10}, ten tree-diagrams contribute to
the final state $u\bar d\mu^-\bar\nu_\mu$, where only three diagrams include a
resonant $W$-boson pair. Similarly, 20 diagrams contribute to the single-$W$
signature $u\bar d e^-\bar\nu_e$.

The consistent treatment of the $W/Z$-boson decay-widths poses 
a further theoretical challenge.  
The {\it Dyson series} allows the resummation of the
self-energy $\Sigma_V$ of the vector boson $V$ to all orders into 
the denominator of the $V$-boson propagator, 
 $(p^2-M_V^2+\Sigma_V(p^2))$.
The complex pole $\mu_V$
 of the propagator defined by
$\mu_V^2-M_V^2+\Sigma_V(\mu_V^2)=0$ provides a gauge invariant 
definition of the mass $M_V$ and width $\Gamma_V$ of the vector bosons,
$\mu_V^2\equiv M_V^2- \mathrm{i} M_V \Gamma_V$.

The Dyson summation of the self-energy includes only a subset of
higher-order diagrams, but neglects other
contributions of the same order.  
A naive application therefore can lead to inconsistencies
such as violations of gauge invariance and unitarity, 
which can can
result in dramatically wrong predictions, in particular in the case of
single-$W$ production  at high energies.\cite{Beenakker:1996kn}
A simple use of a Breit-Wigner propagator with a {\it fixed
  width} 
is sufficient in many leading-order applications, but does not respect
electroweak gauge invariance.  
In  the {\it complex-mass
  scheme}\cite{Denner:2005es}, the replacement $M_V^2\to \mu_V^2$ is
made in the propagator as well as in the Feynman rules,
e.g.\ in the weak-mixing angle  $\cos\theta_w=M_W/M_Z\to
\sqrt{\mu_W^2/\mu_Z^2}$. In this way, algebraic identities among vertices
 and propagators required by gauge invariance are satisfied also for
a finite width.
The \emph{fermion-loop scheme},\cite{Beenakker:1996kn}  
applied in particular to the single-$W$ process at
LEP2,\cite{Accomando:1999zq} uses the fact that diagrams with a closed fermion
loop form a gauge invariant subset of diagrams.
Finally, the {\it double-pole approximation}~(DPA)
consistently splits the NLO corrections into factorizable
corrections to on-shell vector-boson production and decay, and non-factorizable
soft-photon corrections connecting vector-boson production,
propagation and decay. The DPA has been applied to  $W$- and $Z$-boson pair production 
at LEP2.\cite{Jadach:1996hh}

The methods summarized here have been used 
successfully to describe the LEP2 measurements of four-fermion production with a
theoretical accuracy better than $1\%$ for
$W$-pair production and $2$--$5\%$ for the other processes.\cite{Schael:2013ita,Grunewald:2000ju} 

\section{Recent theoretical developments}
\label{sec:recent}

The high accuracy possible at future $\ee$ colliders makes it mandatory to
improve the theoretical predictions of four-fermion cross sections beyond the
level achieved for LEP2.  The $\mw$ measurement from a threshold scan requires
a calculation of the $W$-pair production cross section with a precision of a
few per-mille in the threshold region $\sqrt s\sim 2\mw$, where the accuracy
of the DPA degrades. A complete NLO calculation of charged-current $4$-fermion
production was performed in the complex mass scheme, including loop
corrections to singly- and non-resonant diagrams.\cite{Denner:2005es} The DPA
agrees well with the full $\ee\to 4\mathrm{f}$ calculation for energies
$200\,\gev \lesssim\sqrt{s}\lesssim 500\,\gev$ while the full calculation is
required near threshold $160\,\gev\lesssim \sqrt{s}\lesssim 170\,\gev$, and
for $\sqrt s>500\,\gev$, where off-shell effects become important.  In a
further development, effective-field theory~(EFT) methods have been used for a
dedicated calculation of four-fermion production near the $W$-pair production
threshold.\cite{Beneke:2007zg}
This method has allowed to isolate and compute the
subset of NNLO corrections that is enhanced near threshold due to
Coulomb-photon effects.
Combining these dominant NNLO corrections, which  are of the order of $0.5\%$, with the full NLO
result\cite{Denner:2005es} 
reduces the theoretical uncertainty of the
$\mw$-measurement from a threshold scan to
$\Delta\mw\lesssim\,
3\,\mev$,\cite{Beneke:2007zg} below the ILC precision goal.

At centre-of-mass
energies $\sqrt s \gtrsim 800\,\gev$,
which are particularly relevant for measurements of triple gauge
couplings, higher-order EW corrections are enhanced 
by Sudakov logarithms. 
For $W$-pair production, NNLO  corrections due to NNLL Sudakov logarithms
$\alpha^2\log^m(s/\mw^2)$ with $m=2,3,4$ have been computed\cite{Kuhn:2007ca}
and  are of the order of 
$5\%$ ($15\%)$ for $\sqrt s=1\,\tev$ ($3\,\tev$), so 
they should be taken into account in the second phase of an ILC or at CLIC. 

In addition to these precision calculations, the search for indirect signals
of new physics requires a systematic treatment of deviations from the SM in an
EFT framework, which has recently been applied to study the sensitivity to
anomalous gauge boson couplings in $W$-pair production.\cite{Buchalla:2013wpa}

\section{Outlook}
\label{sec:outlook}

Theoretical methods for higher-order calculations have seen remarkable progress
after the LEP2 era.  The theoretical uncertainty on  charged-current
four-fermion production has been reduced well below the percent level by
a full NLO calculation.\cite{Denner:2005es} The extension of this
calculation to the remaining processes of Fig.~\ref{fig:processes} will be
simplified by recent progress on the automation of EW NLO
calculations\cite{Kallweit:2014xda} and may provide sufficient precision for
future linear $\ee$ colliders, if supplemented with dominant NNLO effects in
special kinematic regions\cite{Beneke:2007zg,Kuhn:2007ca} and an improved
treatment of initial-state radiation.
The precision goals of  future circular $\ee$ colliders may require a full
NNLO calculation of EW corrections, where the current state of the art is
given by $1\to 2$ processes.\cite{Actis:2008ts}
The extension to $2\to 2$ processes such as on-shell vector-boson
pair production is beyond current methods, but may be feasible within several
years. This would provide one of the building blocks of the extension of the
double-pole approximation\cite{Jadach:1996hh} or the EFT
approach\cite{Beneke:2007zg} to NNLO, which in addition requires the
computation of two-loop soft-photon corrections with finite-width
effects. 
 Steps towards a decision on the construction of a future $\ee$
collider would stimulate theoretical developments to meet these challenges.


\begin{thebibliography}{99}

\bibitem{Baer:2013cma} 
  H.~Baer {\it et al.},
  arXiv:1306.6352 [hep-ph];\\
  L.~Linssen {\it et al.}, 
  arXiv:1202.5940 [physics.ins-det];\\
  M.~Bicer {\it et al.} 
  JHEP {\bf 1401}, 164 (2014)
  [arXiv:1308.6176 [hep-ex]];\\
CEPC project \url{http://cepc.ihep.ac.cn/}.
\bibitem{Schael:2013ita} 
  S.~Schael {\it et al.}
  Phys.\ Rept.\  {\bf 532}, 119 (2013)
  [arXiv:1302.3415 [hep-ex]].
\bibitem{Grunewald:2000ju} 
  M.~W.~Grunewald {\it et al.},
  [hep-ph/0005309].
\bibitem{Baak:2013fwa} 
  M.~Baak {\it et al.},
  arXiv:1310.6708 [hep-ph].
%
\bibitem{Beenakker:1996kn} 
  W.~Beenakker {\it et al.}, 
  Nucl.\ Phys.\ B {\bf 500}, 255 (1997)
  [hep-ph/9612260].

\bibitem{Denner:2005es} 
  A.~Denner {\it et al.}, 
  Phys.\ Lett.\ B {\bf 612}, 223 (2005)
  [Erratum-ibid.\ B {\bf 704}, 667 (2011)]
  [hep-ph/0502063];
  Nucl.\ Phys.\ B {\bf 724}, 247 (2005)
  [Erratum-ibid.\ B {\bf 854}, 504 (2012)]
  [hep-ph/0505042].

\bibitem{Accomando:1999zq} 
  E.~Accomando {\it et al.}, 
  Phys.\ Lett.\ B {\bf 479}, 209 (2000)
  [hep-ph/9911489];
  G.~Passarino,
  Nucl.\ Phys.\ B {\bf 578}, 3 (2000)
  [hep-ph/0001212].
\bibitem{Jadach:1996hh} 
  S.~Jadach {\it et al.}, 
  Phys.\ Rev.\ D {\bf 56}, 6939 (1997)
  [hep-ph/9705430];
\\
  W.~Beenakker {\it et al.}, 
  Nucl.\ Phys.\ B {\bf 508}, 17 (1997)
  [hep-ph/9707326];\\
  A.~Denner {\it et al.}, 
  Nucl.\ Phys.\ B {\bf 587}, 67 (2000)
  [hep-ph/0006307];\\
  S.~Jadach {\it et al.}, 
  Phys.\ Rev.\ D {\bf 65}, 093010 (2002)
  [hep-ph/0007012].
\bibitem{Beneke:2007zg} 
  M.~Beneke {\it et al.}, 
  Nucl.\ Phys.\ B {\bf 792}, 89 (2008)
  [arXiv:0707.0773 [hep-ph]];
  S.~Actis {\it et al.}, 
  Nucl.\ Phys.\ B {\bf 807}, 1 (2009)
  [arXiv:0807.0102 [hep-ph]].
\bibitem{Kuhn:2007ca} 
  J.~H.~K\"uhn, F.~Metzler and A.~A.~Penin,
  Nucl.\ Phys.\ B {\bf 795}, 277 (2008)
  [Erratum-ibid.\  {\bf 818}, 135 (2009)]
  [arXiv:0709.4055 [hep-ph]].


\bibitem{Buchalla:2013wpa} 
  G.~Buchalla {\it et al.}, 
  Eur.\ Phys.\ J.\ C {\bf 73}, no. 10, 2589 (2013)
  [arXiv:1302.6481 [hep-ph]];
  J.~D.~Wells and Z.~Zhang,
  arXiv:1507.01594 [hep-ph].
\bibitem{Kallweit:2014xda} 
  S.~Kallweit {\it et al.},
  JHEP {\bf 1504}, 012 (2015)
  [arXiv:1412.5157 [hep-ph]];
  M.~Chiesa, N.~Greiner and F.~Tramontano,
  arXiv:1507.08579 [hep-ph].
\bibitem{Actis:2008ts} 
  S.~Actis {\it et al.}, 
  Nucl.\ Phys.\ B {\bf 811}, 182 (2009)
  [arXiv:0809.3667 [hep-ph]].


\end{thebibliography}
\end{document}